\documentclass[aps,prL,twocolumn,superscriptaddress,showpacs]{revtex4-1}
\usepackage{graphicx}
\usepackage[dvipsnames,usenames]{color}
\usepackage{amsmath} 
\usepackage{amssymb}
\usepackage{bm}
\usepackage{multirow}
\usepackage{wasysym}
\usepackage{subfigure}
\usepackage{verbatim}
\usepackage{marvosym}
\usepackage{latexsym}
\usepackage[dvipsnames,usenames]{color}
\usepackage{float} 
\usepackage{ulem}

\usepackage{xr}
\makeatletter
\newcommand*{\addFileDependency}[1]{
  \typeout{(#1)}
  \@addtofilelist{#1}
  \IfFileExists{#1}{}{\typeout{No file #1.}}
}
\makeatother

\newcommand*{\myexternaldocument}[1]{
    \externaldocument{#1}
    \addFileDependency{#1.tex}
    \addFileDependency{#1.aux}
}

\myexternaldocument{SM}

\begin{document}
\title{Suspensions of small ultra-soft colloids remain liquids in overcrowded conditions}


\author{Nikolaos A. Burger}
\affiliation{Division of Physical Chemistry, Lund University, SE-22100 Lund, Sweden}
\author{Alexander~V.~Petrunin}
\affiliation{Institute of Physical Chemistry, RWTH Aachen University, 52056 Aachen, Germany}
\author{Ann E.~Terry}
\affiliation{MAX IV Laboratory, Lund University, P.O.~Box 118, 22100 Lund Sweden}
\author{R.~Schweins}
\affiliation{Institut Laue-Langevin ILL DS/LSS, 71 Avenue des Martyrs, F-38000 Grenoble, France}
\author{Andrea~Scotti}
\email{andrea.scotti@fkem1.lu.se}
\affiliation{Division of Physical Chemistry, Lund University, SE-22100 Lund, Sweden}

\date{\today}
\begin{abstract}
Concentrated suspensions of small ultra-soft colloids (ultra-low crosslinked microgels) are investigated with scattering and steady shear rheology to capture their equilibrium dynamics. 
The suspensions lack dynamic arrest, although the slow relaxation time $\tau_2$ follows exponential growth with increasing generalized packing fraction, $\zeta$. 
The zero-shear viscosity grows weakly with $\zeta$, and never diverges in contrast to other soft glass formers, e.g.~star-polymers, microgels, green particles. 
Their high compressibility allows these ultra-soft spheres to diffuse even in overcrowded environments.  

\end{abstract}

\maketitle 

 A fundamental understanding of the relationship between the properties of an individual particle (internal architecture, softness, size, and shape) and the flow properties in soft glassy systems remains challenging, debated, and in some cases unexplored. 
 The most studied systems to explore glass transition are suspensions of hard spheres or other soft colloids, such as microgels, star polymers, and micelles \cite{vlassopoulos2001multiarm, brambilla2009probing, Mat09}. 
 The interaction between hard spheres is typically described by a purely repulsive potential.
 The suspensions undergo a glass transition at the volume fraction $\phi = 0.58$, where both viscosity and structural relaxation time diverge super-exponentially at a critical concentration $\phi_0$: $y = y_0 \exp\left( \frac{a\phi}{\phi_0 - \phi} \right)$\cite{van1994glass, berthier2011dynamical}. 
 
 The suspensions of soft, repulsive colloids undergo a liquid–glass transition at higher concentrations \cite{vlassopoulos2014tunable}.  
 This transition is characterized by non-ergodic dynamics and an infinite zero-shear viscosity, forming the so-called yield stress fluids \cite{kamani2021unification, benzi2021stress}. 
 Depending on particle softness, the zero-shear viscosity $\eta_0$ and relaxation time $\tau_2$ may follow a hard-sphere-like behavior orlightly weaker exponential growth~\cite{Van17, Mat09}. 
 The effects of particle compressibility are extensively explored in systems such as multi-arm star polymers \cite{Erw10, vlassopoulos2001multiarm}, microgels \cite{philippe2018glass, gury2025internal}, ring polymers\cite{slimani2014cluster, michieletto2016topologically} and are key to understanding the jamming and glass transition on a fundamental level.
 This is also pivotal for bio-relevant systems such as suspensions of proteins \cite{jawerth2020protein} or green nanoparticles \cite{shamana2018unusual} that also form soft glasses. 
 
 Here, we investigate the equilibrium dynamics of microgel suspensions combining rheology, dynamic light scattering (DLS), and small-angle neutron and X-ray scattering (SANS and SAXS, respectively), to elucidate the effect of particle softness and structure on the internal dynamics both at rest and under flow.
 To unravel the effects of particle size and softness, we study poly(\textit{N}-isopropylacrylamide) (pNIPAM) microgels synthesized both without the addition of crosslinker (ultra-low crosslinked, ULC, with hydrodynamic radius $R_H = 114 \pm2$~nm) and with 3.7~mol\% crosslinker agent (regularly crosslinked, RC, $R_H = 107\pm1$~nm), see Sec~\ref{sec:synt}.
 Indeed one can finely tune  the particle bulk modulus for microgels simply changing the amount of crosslinker agent used during the synthesis \cite{Hou22, Hof24}. 
 
 We find that the values of the slow relaxation time $\tau_2$ of the ULC microgel suspensions follows an exponential growth and $\tau_2$ diverges with increasing the generalized packing fraction $\zeta$, and reveals a transition at $( \zeta \approx 2.65)$, which is significantly higher compared to the value for hard-sphere systems and harder microgels like the RC. 
 This shift is attributed to particle compressibility and osmotic deswelling \cite{Sco16}. 
 In contrast to other soft-colloidal systems however, the evolution of zero-shear viscosity with $\zeta$ for the ULC microgel suspensions exhibits no divergence and fluid-like dynamics are maintained even in highly overcrowded environments. 

To confirm that the synthesis of the ULC microgels effectively led to particle formation and not simply to high molecular weight polymer, we used small-angle x-ray scattering, SAXS, on the coSAXS beamline at MAX-IV (Sec.~\ref{sec:SAXS}) to access their form factors.
Indeed the form factor of a linear polymer will show a decay of the intensity with increasing $q$, ($I\sim q^{-1.7, -2}$, depending on the solvent quality), while the form factor of a spherical particle will show oscillation \cite{Gla82} and ($I\sim q^{-4}$, spheres).
Furthermore, from the low-$q$ region, one can extract the particle radius of gyration $R_g$.
Again, in the case of linear polymer in good solvent the ratio $R_G/R_H$ is larger than one \cite{Wan99}, while for spherical particles is $\lesssim (3/5)^{0.5}$.

\begin{figure}[htbp!]
    \centering
    \includegraphics[trim=2.2cm 0cm 0cm 0cm, clip, width=0.49\textwidth]{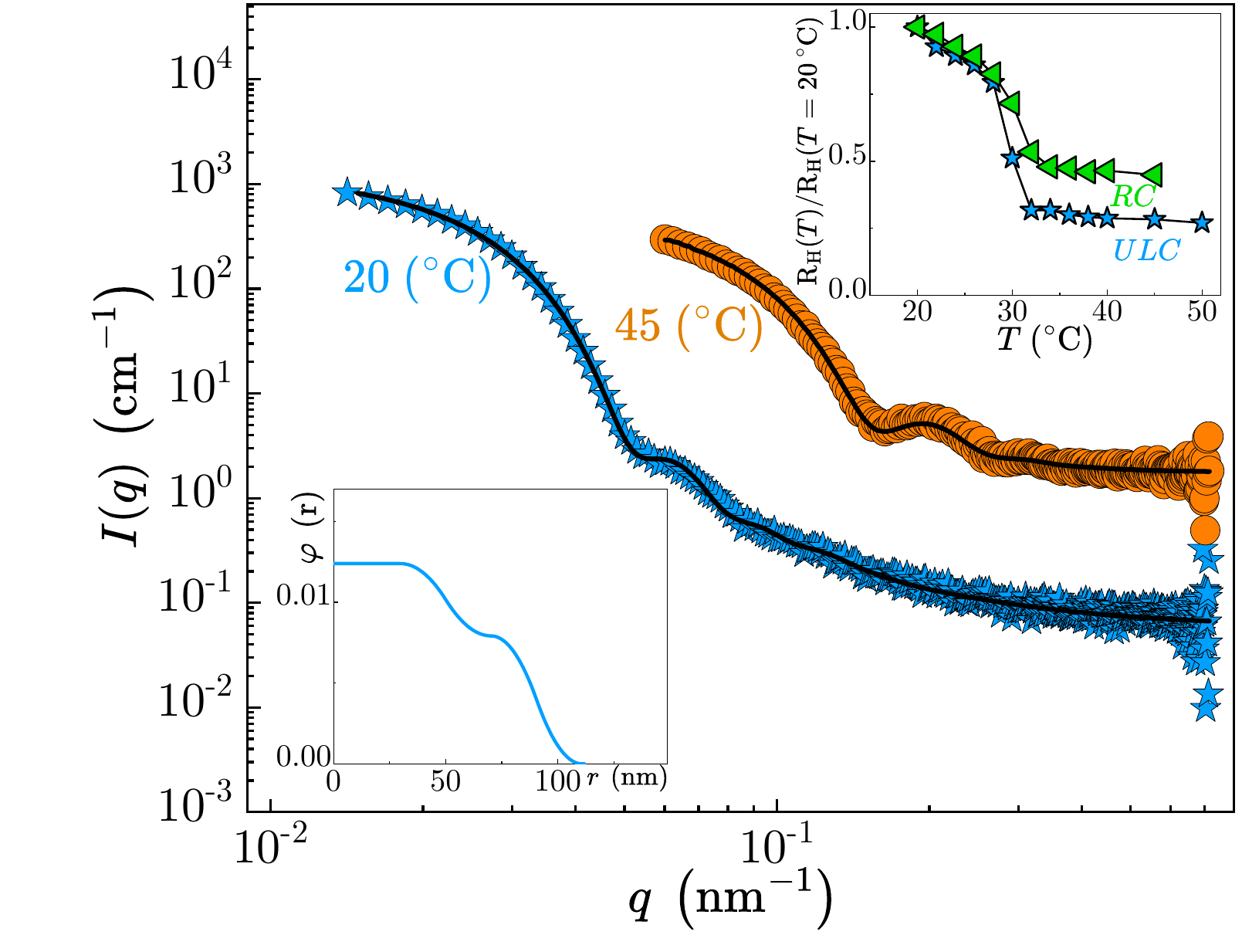}
    \caption{Small-angle X-ray scattering intensities $I(q)$ versus scattering vector $q$ of ULC microgels measured at 20~$^\circ$C (stars) and 45~$^\circ$C (circles).
    The top curve is shifted vertically by a factor of $10^2$ for clarity.
    Solid lines: fits of the data with a with a core-fuzzy-shell model \cite{Ber06}. \textit{Bottom inset:} Relative polymer radial distribution as obtained from the fit of the form factor. 
    \textit{Top inset:} Hydrodynamic radius normalized with the hydrodynamic radius at 20~$^\circ$C, versus temperature $T$, for the 3.7~mol\% (triangles) and ultra-low crosslinked microgels (stars).}
    \label{fig:PqSAXS}
\end{figure}

Fig.~\ref{fig:PqSAXS} shows the SAXS intensities $I(q)$ of dilute suspensions of ULC microgels measured in the swollen state at $T = 20~^\circ$C (stars) and in the collapsed state at $45~^\circ$C (circles).
In both cases, oscillations are clearly visible in the region $4\cdot10^{-2}$~nm$^{-1} \lesssim q \lesssim 8\cdot10^{-2}$~nm$^{-1}$ and $10^{-1}$~nm$^{-1} \lesssim q \lesssim 3\cdot10^{-1}$~nm$^{-1}$ at 20 and 40~$^\circ$C, respectively.
Therefore, the synthesis without the addition of crosslinker yields particles.

The black solid lines represent the fits of the data with a core-fuzzy-shell model \cite{Ber06} which accounts for the possibility of having a core with a different scattering length density with respect to the less dense shell (Sec.~\ref{sec:mod}).
This model was necessary since fitting the data with the common fuzzy-sphere models \cite{Sti04}, leads to unreliably high values for the total particle radius compared to the $R_H$ obtained from DLS.
In contrast, the RC microgels show the expected fuzzy-sphere structure (see Fig.~\ref{fig:PqSAXSR5}).
For the ULC microgel we obtained $M_w = (4.5 \pm 0.3) \cdot 10^7$~g/mol, while for the RC microgels $ (1.5 \pm 0.1) \cdot 10^8$~g/mol.
The values of the fitting parameters are reported in Table~\ref{tab:SAXSPq}: the total radius of the ULC microgels is $(111 \pm 4)$~nm at 20~$^\circ$C and $(27\pm2)$~nm at 45~$^\circ$C.
The bottom inset of Fig~\ref{fig:PqSAXS} shows the radial distribution as obtained from the fit of the data at 20~$^\circ$C, which reveals a inhomogeneous distribution of polymer within the ULC particles with a denser core, $(30\pm1)~$nm, and a poorly polymer rich shell ($31~\text{nm}<R<111$~nm).
Such a radial distribution indicates that the ULC microgels have a denser core surrounded by a scarcely crosslinked shell with very low polymer density.
One can obtain the particle radius of gyration $R_G$ fitting the $I(q)$  in the $q$-region where $qR_G < 1.2$, Fig.~\ref{fig:guinier}. 
For a homogeneous sphere, the ratio $R_G/R_H$ equals $(3/5)^{0.5} \approx 0.77$.
For the ULC we obtain $R_G / R_H \sim 0.68 \pm 0.07$ at $20~^\circ$C, and $R_G / R_H \sim 0.77 \pm 0.065$ at $45~^\circ$C.

The radial distributions $\varphi(R)$ also show that the ULC microgels contain much less polymer within their volume compared to the RC.
This is confirmed by the molecular weight of the particle computed combining viscosimetry and DLS measurements on very dilute suspensions (see Secs.~\ref{sec:visc} and \ref{sec:DLS}). 

The upper inset of Fig.~\ref{fig:PqSAXS} shows the deswelling ratio  $R_H(T =50~^\circ\text{C})/R_H(T =20~^\circ\text{C})$ vs.~temperature for dilute suspensions of ULC (stars) and RC (triangles). 
For RC, the deswelling ratio is $0.46 \pm 0.009$, consistent with the values reported in the literature for similar crosslinked microgels \cite{Sco22_review}. 
In contrast, for the ULC microgels  $R_H(T =50~^\circ\text{C})/R_H(T =20~^\circ\text{C})= 0.27 \pm 0.01$. 
This value is significantly smaller than the one previously reported \cite{Sco22_review}, even compared to recently introduced ultra-soft star-like microgels~\cite{Bal25} and indicate a very soft nature of these particles \cite{Sco22_review}.

To quantify the softness of the microgels used here, we applied a method recently introduced to measure the bulk modulus $K$ of particles with radii below few hundreds of nm which is based on small-angle neutron scattering and contrast variation. 
The measurements were performed on D11 at the Institut Laue Langevin \cite{Hou22} ( Sec.~\ref{sec:SANS}).
The bulk modulus of microgels does not remain constant but increases with compression, indicating that it is harder and harder to further compress the particle.
The starting value of $K$ for the ULC is ($3.5\pm0.5$)~kPa, comparable with what was reported before for larger ULC microgels \cite{Hou22, Hof24}, similar to the value of $K$ recently determined for ultra-soft star-like microgels~\cite{Bal25}.
Upon compression, the value of $K$ increases and reaches a maximum of $(56\pm1)$~kPa when the volume of the particles decreases of more than 60\% with respect to the original value, Fig.~\ref{fig:K}. 
For RC microgels, the initial value of $K$ is three times larger than what is measured for the ULC. The values also increase steeply and reach a value above 100~kPa once the compression exceeds 60\% of the original volume, in line with what was previously reported in the literature \cite{Hou22}. 

The combination of viscosimetry, DLS, SAXS and SANS measurements confirms that the synthesis without the addition of any crosslinker agent led to small ultra-soft microgels and not simply to polymeric chains.
However, due to their smaller size, they present a peculiar radial distribution characterized by a denser small core surrounded by a shell of poorly crosslinked polymer and dangling chains. Simply tuning the size of the particle and performing a standard precipitation polymerization synthesis, one can obtain ultra-soft particles with a very soft and hairy shell similar to the one obtained using a crosslinker that reacts faster than the usual $N,N'$-methylenebisacrylamide.

We now proceed to analyze the dynamics in concentrated suspensions of these small ULC microgels.
The samples concentration is expressed using the generalized volume fraction $\zeta$ that is the fraction of volume occupied by the microgels in the suspensions assuming they maintain their swollen radius.
Since microgels respond to crowding by deswelling, faceting and interpenetrating each other \cite{Sco16, Con17, Sco22_review}, the values of $\zeta$ can exceed the unity.
The conversion constants to pass from weight fraction of polymer in solution to $\zeta$ were obtained from viscosimetry measurements for both the ULC and RC microgels \cite{Sco21VF} (Sec.~\ref{sec:visc}).
Notably, the suspension of ULC microgels remained transparent even at high $\zeta$ (Fig.~\ref{fig:photo}), allowing the use of conventional DLS to probe dynamics across a wide range of wavevectors $q$ and packing fractions $\zeta$ \cite{Pus87}. 

\begin{figure}[htbp!]
    \centering
    \includegraphics[trim=5.5cm 0.25cm 6cm 0cm, clip, width=0.49\textwidth]{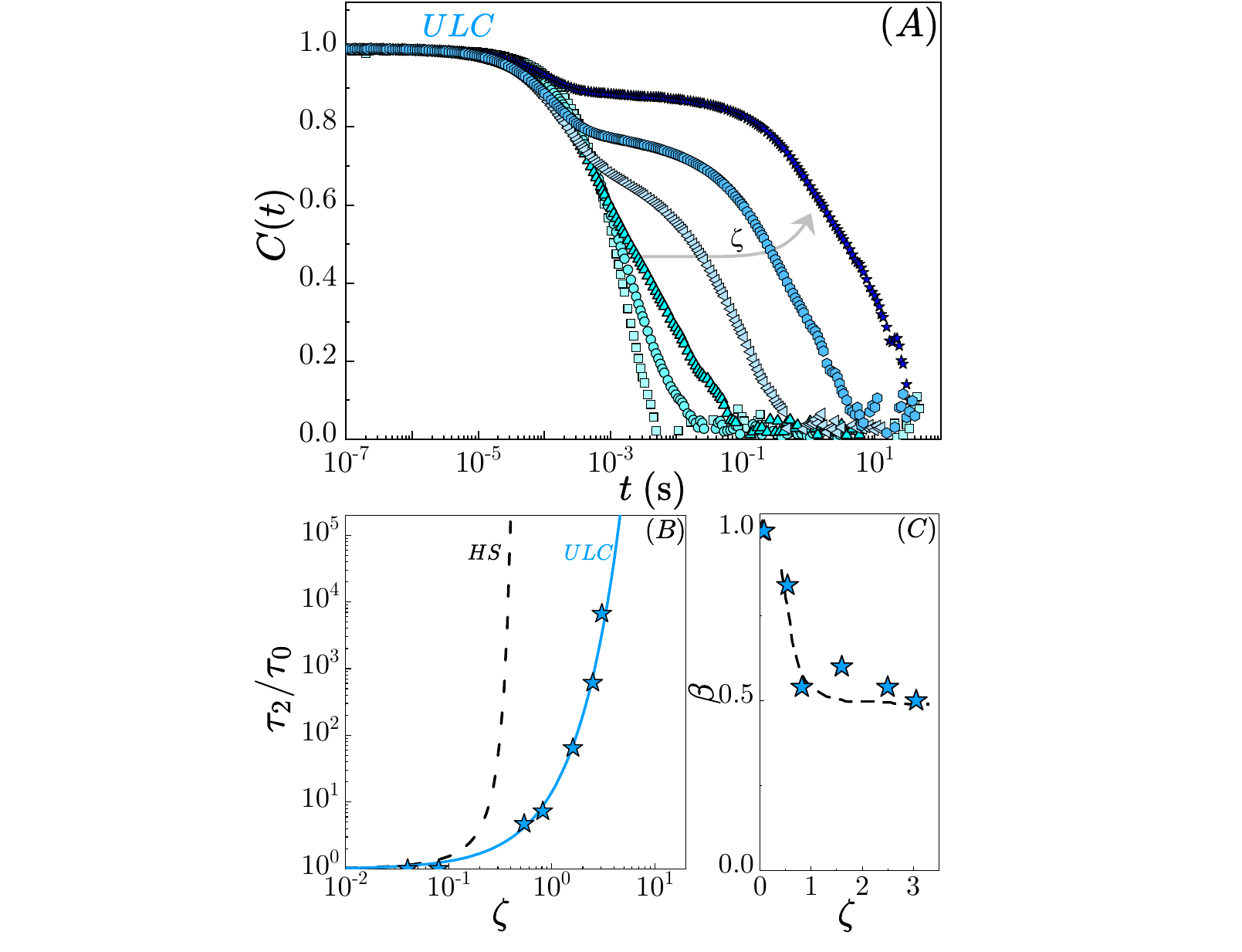}
    \caption{(A) Intermediate scattering function (ISF) C(t) for ULC microgels at $q = 1.79 \times 10^{-2}$~nm$^{-1}$, where the generalized packing fraction $\zeta$ ranges from $0.04\pm 0.01$ (squares), $0.08\pm 0.01$, $0.54\pm 0.01$, $0.82\pm 0.01$, $1.6\pm 0.01$, $2.5\pm 0.01$ to $3.06\pm 0.01$ (stars), respectively.
(B) Normalized relaxation time (with respect to the relaxation time, $\tau_0$ at $\zeta = (0.04\pm0.01$) of ULC particles versus $\zeta$. 
The dashed line corresponds to hard-sphere (HS) particles with a fluid-glass transition at $\zeta_g = 0.58$. 
(C) Stretched exponent $\beta$ versus $\zeta$ for ULC particles.}
    \label{fig:DLS}
\end{figure}

Fig.~\ref{fig:DLS}A shows the intermediate scattering function $C(t)$ for ULC particles at different $\zeta$. 
In the dilute regime, $C(t)$ is described by a single exponential decay: $C(t) = A_0 \exp(-t/\tau_0)$. 
As $\zeta$ increases, a second relaxation process emerges at intermediate packing fractions. 
This secondary process progressively slows down with further increase in $\zeta$. 
At high packing fractions, the intermediate scattering function is well described by: $C(t) = A_1 \exp(-t/\tau_1) + A_2 \exp\left[ -\left(t/\tau_2\right)^\beta \right]$. 
It is important to note that both relaxation processes are characterized by diffusive motion. 
The derived relaxation time $\tau_2$ is normalized by the relaxation time in the dilute regime, $\tau_0$ and plotted as a function of $\zeta$ in Fig.~\ref{fig:DLS}B. 
For hard spheres (HS), the evolution of $\tau_2$ with $\zeta$ deviates super-exponentially and follows a Vogel–Fulcher–Tammann (VFT)-like growth: $ \tau_2 = \tau_0 \exp\left( \frac{a \zeta}{\zeta_0 - \zeta} \right)$, typical of repulsive glasses (dashed line in Fig.~\ref{fig:DLS}(B))~\cite{berthier2011dynamical}. 

In contrast, our ULC particles show a much weaker dependence on $\zeta$, and the evolution of $\tau_2$ is well captured by an exponential-like growth: $\tau_2 = \tau_0 \exp(a \zeta)$, solid line in Fig.~\ref{fig:DLS}(B). 
This deviation remains stronger than what was observed in the literature for polymer solutions \cite{rubinstein2003polymer}. 
We also analyze the variation of the stretched exponent $\beta$ as a function of $\zeta$, shown in Fig.~\ref{fig:DLS}C. 
The exponent $\beta$ initially decreases from 1 with increasing $\zeta$, until it reaches a plateau at $\beta \sim 0.5$ at higher packing fractions. 
This behavior is reminiscent of what has been reported in molecular glasses and other soft glassy systems~\cite{Mat09}.

\begin{figure}[htbp!]
    \centering
    \includegraphics[trim=5cm 2.9cm 0.2cm 0.6cm, clip, width=0.49\textwidth]{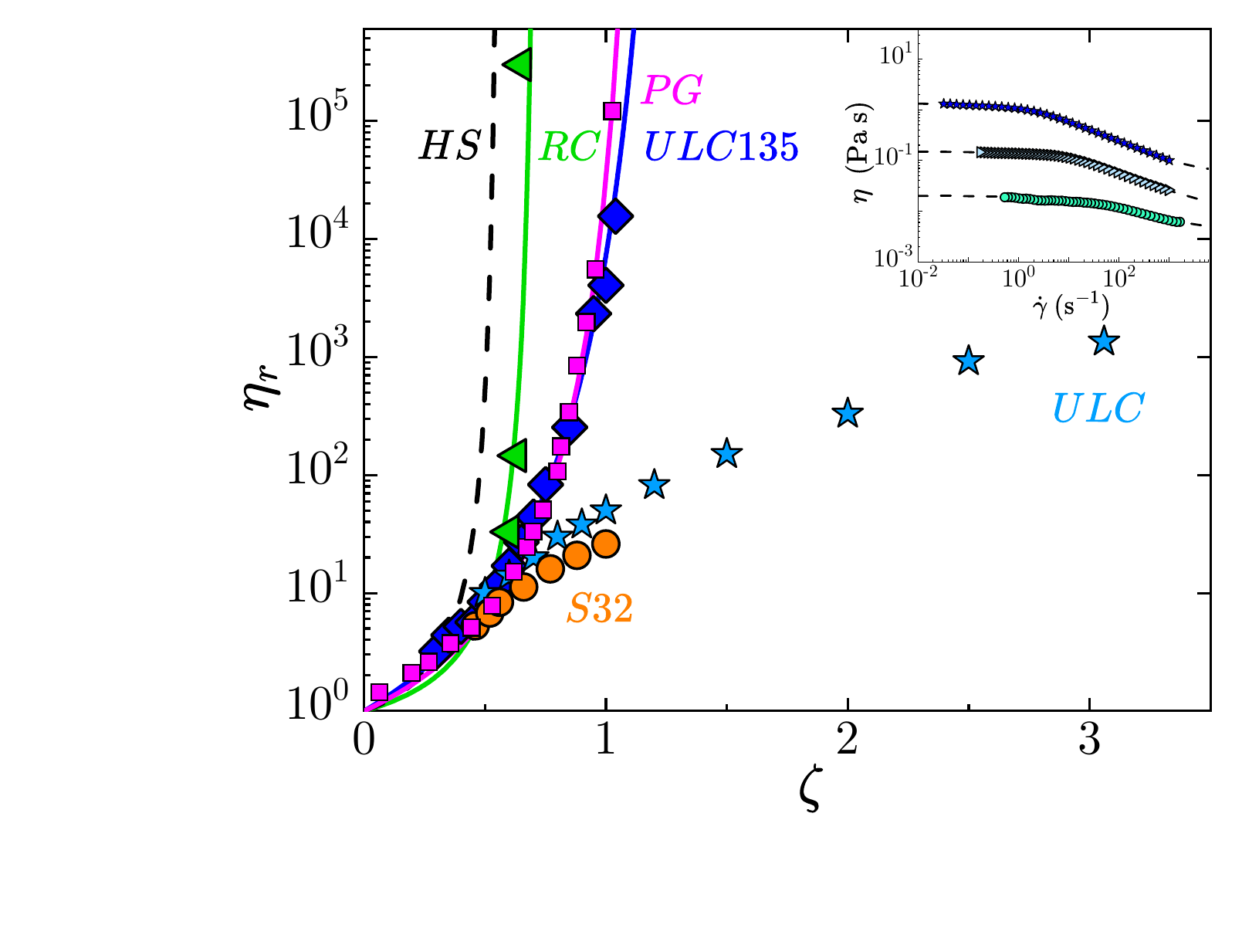}
    \caption{Relative viscosity, $\eta_r$, versus generalized volume fraction, $\zeta$, for: 3.7~mol\% crosslinked microgels, RC (triangles), ULC microgels, ULC (stars), larger ULC microgels, ULC135 (diamonds) derived from~\cite{Sco20_flow},low-arm star, S32, (circles) derived from ~\cite{vlassopoulos2001spherical} and phytoglycogen nanoparticles, PG (squares) derived from ~\cite{shamana2018unusual}. 
    The black dashed line represents the fitting for hard-sphere (HS) particles. The solid lines represent fitting of the data with the VFT model.
    {Inset:} Steady shear viscosity, $\eta$, versus shear strain rate, $\dot{\gamma}$, of ULC suspensions with packing fractions from bottom to top: $\zeta = 0.70\pm 0.01$ (circles), $1.50\pm 0.01$ (triangles), and $3.06\pm 0.01$ (stars). Dashed lines: fits of the data with \label{eq:Cross-equation}.}
    \label{fig:visc_main}
\end{figure}

We now focus on the flow properties of the ULC microgels as a function of $\zeta$.
The inset of Fig.~\ref{fig:visc_main} shows representative steady shear measurements for the viscosity $\eta$ of ULC suspensions at different $\zeta$. The course of $\eta$ has the expected shear thinning behavior, and it is characterized by a plateau in the viscosity $\eta_0$ at low shear rates $\dot\gamma$.
The dashed lines are fits of the data with Eq.~\ref{eq:Cross-equation} used to obtain the value of $\eta_0$ (Sec.~\ref{sec:rheo}).

Fig.~\ref{fig:visc_main} shows the evolution of the relative viscosities $\eta_r$, i.e.~the ratio between $\eta_0$ and the solvent viscosity (water) with $\zeta$.
We compare the behavior of ULC (stars) and RC (triangles) microgels, HS (dashed line), and larger ULC particle with $R_H = (138.3\pm0.6)$~nm (diamonds).  
For suspensions of RC microgels, a divergence of $\eta_r$ with $\zeta$ is observed. 
It follows a Vogel–Fulcher–Tammann (VFT)-like function, where $\eta_r = \eta_0 \exp\left( \frac{a\zeta}{\zeta_0 - \zeta} \right)$. 
The only difference is that $\zeta_0 \simeq 0.7$ instead then being equal to 0.58. 

Surprisingly, the ULC particles follow a much weaker dependence and no divergence of $\eta_r$ is observed (stars). 
The growth of the relative viscosity with $\zeta$ is similar to the one reported for low-arm (32-arm) star polymer solutions (circles) and much weaker compared to higher-arm star (128-arm), micelle suspensions, and phytoglycogen nanoparticles~\cite{Erw10, merlet2010swelling, roovers1994, shamana2018unusual}. 
We note that phytoglycogen nanoparticles with a size similar to ULC particles but a larger bulk modulus (15~kPa) show a divergence of $\eta_r$, albeit at very high $\zeta$.
Similarly, larger ULC particles that have comparable values of $K$ to the one of the smaller ULC microgels used here, also present a divergence of $\eta_r$ at $\zeta_0 \simeq 1.45$.
These two facts indicate that the combination of the small size and low bulk modulus of the ULC microgels studied here is the key to suppress the divergence of $\eta_r$.

\begin{figure}[htbp!]
    \centering
   \includegraphics[trim=1cm 1cm 3cm 2cm, clip, width=0.49\textwidth]{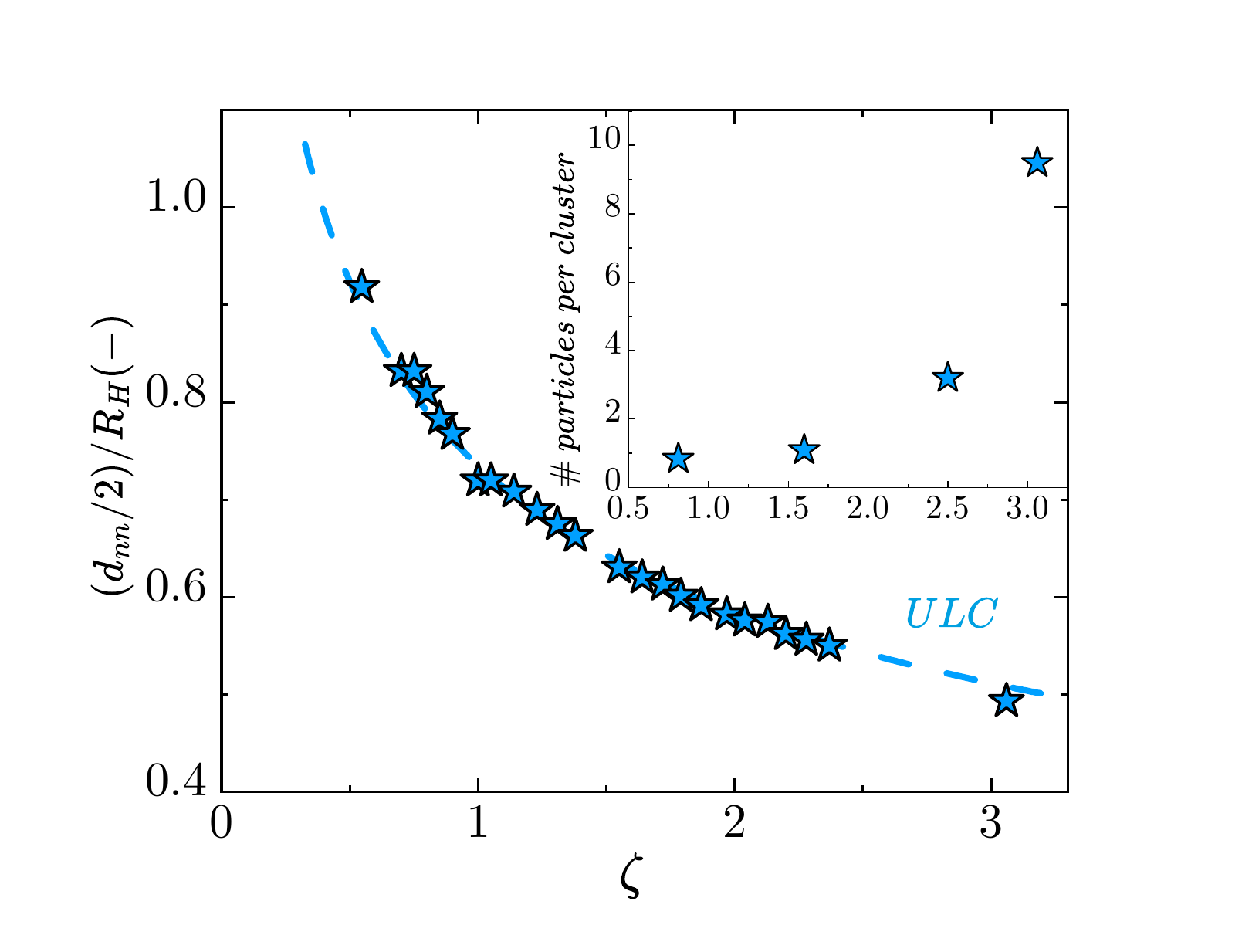}
    \caption{Nearest-neighbor distance, $d_{nn}$, determined from SAXS as a function of generalized volume fraction, $\zeta$, for ULC particles (stars). 
    The dashed black line is a fit of the data with $d_{nn} = z \zeta^{-1/3}$. 
    \textit{Inset:} Apparent number of particles per cluster as a function of $\zeta$ as determined by DLS.}
    \label{fig:dnn}
\end{figure}

To further probe the particle-to-particle arrangement within our samples, we performed SAXS on concentrated samples (Fig.~\ref{fig:SAXS_ULC}).
The position of the structural peak $q_{max}$ can be used to obtain the value of nearest neighbor distance between the particles with increasing $\zeta$: $d_{nn} = 2\pi/q_{max}$.
The values of $d_{nn}$ for the ULC microgel solutions are shown as a function of $\zeta$ in Fig.~\ref{fig:dnn} (stars).
The distance between the microgels decreases with increasing $\zeta$ passing from particles being at a distance comparable with their diameter to a distance that is comparable to the particle radius.
This means that the particle are strongly compressed with increasing $\zeta$.
The dashed line represents a fit of the data with a power law $\propto \zeta^{-1/3}$, which is proper of particles interacting via a soft repulsive potential \cite{Sco22_review}.
Such a decrease in size is due to the combination of isotropic deswelling due to osmotic pressure effects \cite{Sco16, Zho23, Hof22} and faceting \cite{Con17, Con19, Pel15} of the particles.
One can expect a strong deswelling/deformation of the portion of microgels with a very low polymer density ( 31~nm$<R<111$~nm, bottom inset of Fig.~\ref{fig:PqSAXS}). 
To properly assess the size of an individual particle in concentrated suspensions, one should use small-angle neutron scattering with contrast variation \cite{Sco16, Sco21VF, Noj18}, however, in first approximation $d_{nn}/2$ can give us an idea of the average particle size. 

Now, the slow relaxation process measured by DLS reflects the average time a particle needs to escape from a cage formed by neighboring particles.
From the measured slow relaxation time, one can calculate an effective hydrodynamic radius: $R_{H, eff} = \frac{k_B T}{6 \pi D \eta_0}$, with~$\eta_{0}$ the zero-shear viscosity. 
$R_{H, eff}$ can be considered as the average size of a dynamic cluster composed of a specific number of neighboring particles. It follows that the average number of particles per cluster is $R_{H, eff}/(d_{nn}/2)$. 
This quantity is plotted versus $\zeta$ in the inset in Fig.~\ref{fig:dnn}. 
At low $\zeta$, $R_{H, eff}/(d_{nn}/2)=1$, indicating that the slow relaxation process in DLS reflects the motion of individual particles in a more viscous suspension.  
 For $\zeta>1.5$, the increase of $R_{H, eff}/(d_{nn}/2)$ reflects the cluster formation. 
 The number of particles per cluster further increases with $\zeta$ until the clusters reach a total average size of $\approx 500$~nm ($\approx$ 10 particles per cluster). 

Given the absence of aging, the picture that emerges is that our system is characterized by rapid rearrangement of the particles and is consistent with the formation of diffusive and dynamic clusters.  
The ability of ultra-soft and small colloids to deswell and facet makes it easier for them to squeeze through small volumes and diffuse even in highly overcrowded environments.
In particular, simply reducing the size of the microgels makes their mesh-size $\xi$ comparable with the total particle dimension (Tab.~\ref{tab:SAXSPq}, $\xi/R \lesssim 10\%$ versus a usual 15-20\%)
This explains why we see such behavior only for very small ULC microgels.
In conclusion, this letter has shown that reducing the size of microgels synthesized by precipitation polymerization without the addition of a crosslinker make these particles able to escape their neighbors, even in highly overcrowded environments.
What is reported here contrasts with what was observed for other hard and soft colloids. 
As the colloidal glass transition approaches, the free volume vanishes, and particles are forced into contact without being able to move out of their cage at reasonable timescales.
A key aspect here is the comparability between the total size of the particles and their mesh size, which leads to a low value of $K$, even when the particles are strongly compressed.
Due to the large compressibility, these microgels continue to shrink while their concentration increases.
Thanks to this mechanism, the suspensions show a mild increase of internal relaxation times and the viscosity never diverges with increasing $\zeta$.
In other words, our system is reminiscent of molecular liquids and approaches the glass transition in a smooth way without showing dynamic arrest.
Further studies, combining molecular dynamics simulation and experimental data, are needed to further probe the architecture of these particles and, for instance, determine if their shell is crosslinked or mainly composed of dangling chains.
In the second case, they would be similar to thermoresponsive star-like particles \cite{Bal25} but obtained simply by reducing their total size. 
Additional study will also shed light on the influence of the ratio between particle mesh-size and radius on the overall deformability of the microgel and on its effect on making it escape crowding.

\begin{acknowledgments}
AS acknowledge financial support from the Knut and Alice Wallenberg Foundation (Wallenberg Academy Fellows) and from the Swedish Research Council (Research Grant 2024-04178). 
Small-angle neutron scattering measurement have been performed at the Institute Laue Langevin under proposal numbers 9-11-2205 and 9-11-2217 using the instrument D11 and the data are available at http://doi.ill.fr/10.5291/ILL-DATA.9-11-2205 and
http://doi.ill.fr/10.5291/ILL-DATA.9-11-2217. 
Small-angle x-ray scattering experiment have been conducted at the CoSAXS beamline at the MAX IV laboratory (Lund, Sweden) under the proposals 20250233 and 20250032.
The deuterated polymer used here have been provided by the deuteration service at the Forschungszentrum Jülich.
\end{acknowledgments}

\bibliographystyle{apsrev4-1}
\bibliography{References}

\end{document}


\title{\emph{Supplementary Materials:} Suspensions of small and ultra-soft colloids remain liquids in overcrowded conditions}


\author{Nikolaos A. Burger}
\affiliation{Division of Physical Chemistry, Lund University, SE-22100 Lund, Sweden}
\author{Alexander~V.~Petrunin}
\affiliation{Institute of Physical Chemistry, RWTH Aachen University, 52056 Aachen, Germany}
\author{Ann E.~Terry}
\affiliation{MAX IV Laboratory, Lund University, P.O.~Box 118, 22100 Lund Sweden}
\author{R.~Schweins}
\affiliation{Institut Laue-Langevin ILL DS/LSS, 71 Avenue des Martyrs, F-38000 Grenoble, France}
\author{Andrea~Scotti}
\email{andrea.scotti@fkem1.lu.se}
\affiliation{Division of Physical Chemistry, Lund University, SE-22100 Lund, Sweden}

\date{\today}
\begin{abstract}

\end{abstract}

\maketitle 


\section{Synthesis}\label{sec:synt}

For the synthesis of ultralow-crosslinked (ULC) microgels, NIPAM (2.3769~g, 0.07~M) and sodium dodecyl sulfate (0.1730~g, 2~mM), SDS, were dissolved in 295~mL double-distilled H$_2$O that was filtered through a 0.2 $\mu$m syringe filter. 
The solution was brought to 70$^\circ$C in a three-neck flask. 
Potassium persulfate (0.1265 g, 1.56~mM), KPS, was dissolved in 5~mL of the same double-distilled H$_2$O. 
Both solutions were bubbled with N$_2$ for 45 min, after which the reaction was started by adding the KPS into the flask. 
The reaction mixture was kept at 70$^\circ$C under N$_2$ flow for 4.5~h. 
The microgels were purified by five-fold ultra-centrifugation at 50,000 rpm followed by redispersion in fresh double-distilled water. 
Freeze-drying was used for storage.

For the synthesis of RC microgels, NIPAM (5.1607~g, 0.152~M),  $N,N'$-methylenebisacrylamide) (0.2714~g, 0.006 M), BIS, and SDS (0.1081 g, 1.25~mM) were dissolved in 295 mL double-distilled H$_2$O that was filtered through a 0.2~µm syringe filter. 
The solution was brought to 70$^\circ$C in a large flask. KPS (0.1265 g, 1.56 mM) was dissolved in 5~mL of the same double-distilled H$_2$O. 
Both solutions were bubbled with N2 for 60 min, after which the reaction was started by adding the KPS into the flask. The reaction mixture was kept at 70$^\circ$C under N$_2$ flow for 4 h. The microgels were purified by five-fold ultra-centrifugation at 40,000 rpm followed by redispersion in fresh double-distilled water. Freeze-drying was used for storage.

\section{Small-angle x-ray scattering}\label{sec:SAXS}

The SAXS intensities for static samples, contained in quartz capillaries (Hilgenberg) with 1.5 mm diameter and wall thickness of 0.1 mm have been measured at the CoSAXS beamline at the 3 GeV ring of the MAX-IV Laboratory (Lund, Sweden).
The $q$-range of interest between 7$\cdot10^{-3}$ and 0.7~nm$^{-1}$ was covered on CoSAXS using a sample-to-detector distance of 14.2~m with x-ray beam energy E~=~12.4~keV.
The instrument is equipped with an Eiger2 4M SAXS detector with a pixel size of 75~$\mu$m x 75~$\mu$m. 
A python-based code was used to convert the 2D images to 1D profiles.

\begin{table*}
    \centering
    \begin{tabular}{c|c|c|c|c|c|c|c|c|c}
        Sample & T~($^\circ$C)  & $R_H$~(nm) & $R$~(nm)  & $W_{core}$~(nm) & $2\sigma_{in}$~(nm) & $W_{sh}$~(nm)  & $\sigma_{out}$~(nm) & $\xi$~(nm) & $p$~(\%) \\
        \hline
        \hline
         ULC& 20& $114 \pm2$ & $111 \pm4$ & $30\pm1$ &$41\pm2$  & -  & $40\pm1$ & $12\pm1$ &$12\pm1$\\
         RC&  20& $107\pm1$&  $107\pm3$ & $12\pm1$ & $95\pm2$ - & - & - & $8\pm1$ &$8.3\pm0.9$ \\
         ULC& 45& $ 32.6\pm0.5$ & $27\pm2$ & $27\pm2$ & - & - & - & $5.3\pm0.9$ &$12.1\pm0.8$ \\
         RC&  45& $ 50.4\pm0.3$& $ 45\pm2$ & $ 45\pm2$ & -  & - & - & $4.0\pm0.6$ & $8.7\pm0.8$\\
         \hline
    \end{tabular}
    \caption{Values of the fit parameters from the analysis of the SAXS $P(q)$ for the ULC and RC microgels using the model in Eqs.~\ref{eq:mod1} and~\ref{eq:mod2}.}
    \label{tab:SAXSPq}
\end{table*}

\begin{figure}[htbp!]
    \centering
    \includegraphics[width=0.49\textwidth]{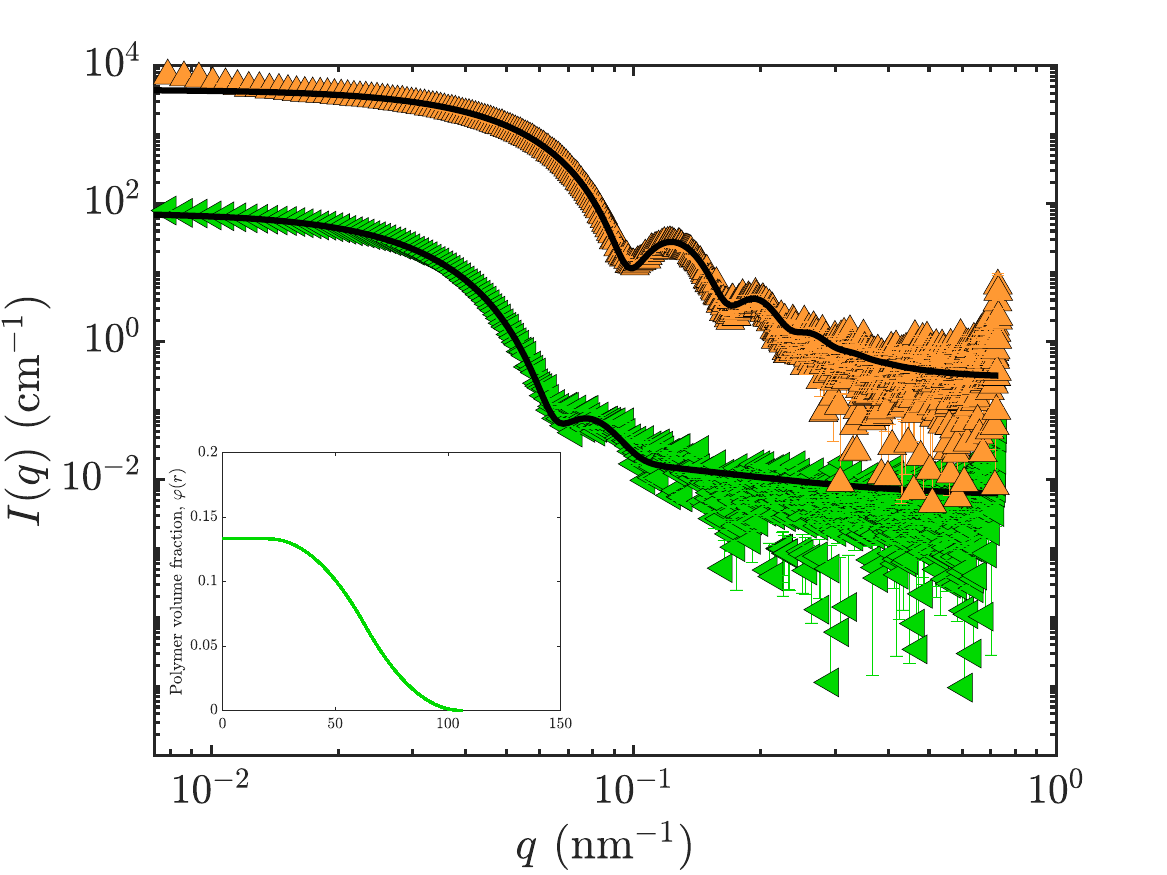}
    \caption{Small-angle X-ray scattering intensities $I(q)$ versus scattering vector $q$ of RC microgels measured at 20~$^\circ$C (left-side triangles) and 45~$^\circ$C (up-side triangles).
    The top curve is shifted vertically by a factor of $10^2$ for clarity.
    Solid lines: fits of the data with a fuzzy-sphere model \cite{Sti04}. 
    \textit{Inset:} Relative polymer radial distribution as obtained from the fit of the form factor. 
    }
    \label{fig:PqSAXSR5}
\end{figure}

\begin{figure}[htbp!]
    \centering
    \includegraphics[trim= 0cm 6cm 0cm 3cm, clip, width=0.49\textwidth]{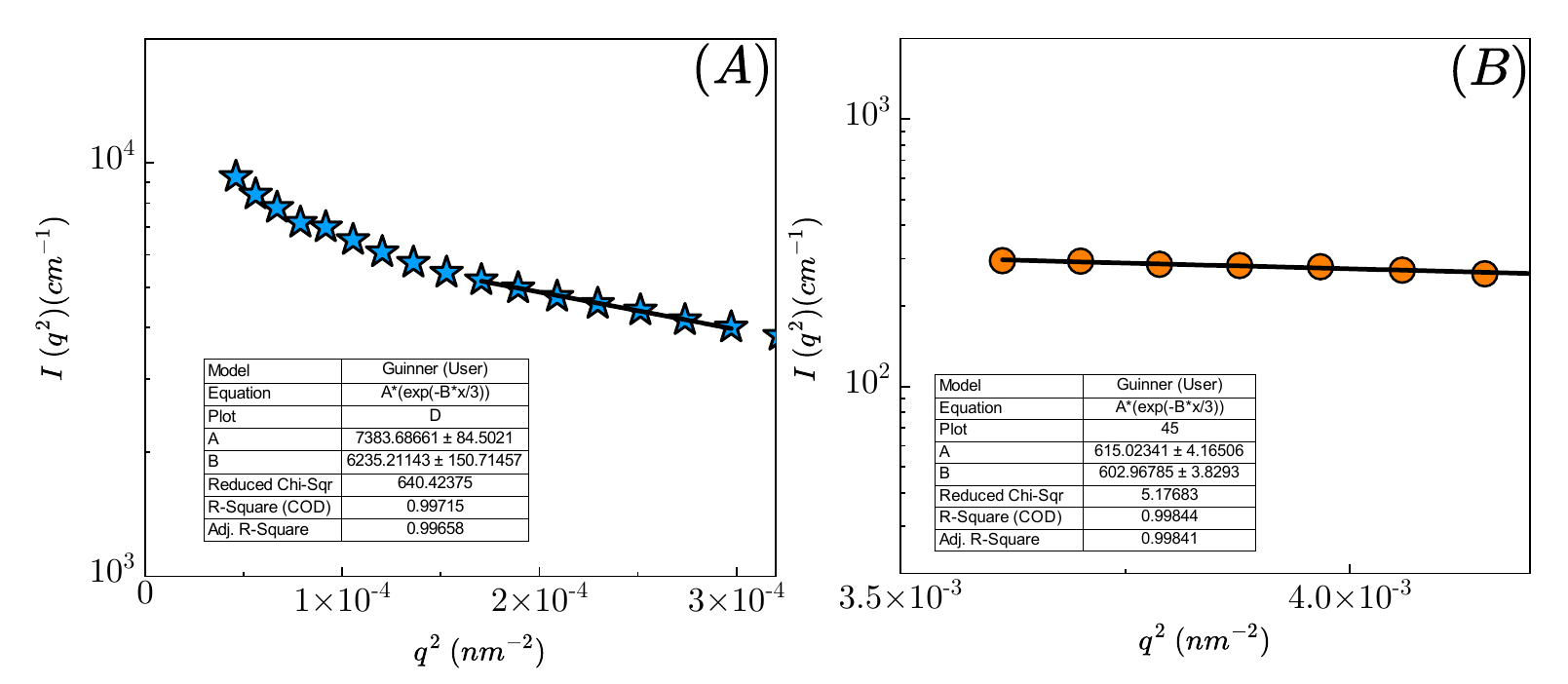}
    \caption{Small-angle X-ray scattering intensities $I(q)$ versus scattering vector $q$ of ULC microgels measured at (A) 20~$^\circ$C (stars) and (B) 45~$^\circ$C (spheres). In plot (A), we didn't count for the initial slope due to possible particle aggregation. 
    Solid lines: fits of the data with a Guinier's model at low $q$-regime \cite{putnam2016guinier}. 
    }
    \label{fig:guinier}
\end{figure}

Fig.~\ref{fig:PqSAXSR5} shows the data relative to SAXS measurements in diluted suspensions of RC microgels at 20~$^\circ$C (left-side triangles) and 45~$^\circ$C (up-side triangles).
The solid line is a fit with the simple fuzzy-sphere model since in contrast to the case of the ULC microgel in Fig.~\ref{fig:PqSAXS}, the use of this simpler model leads to both a good fit and a value of the total radius comparable to the hydrodynamic radius measure with DLS, see Tab.~\ref{tab:SAXSPq}.
We note that, when we use the model in Eqs.~\ref{eq:mod1} and \ref{eq:mod2}, we obtain the same radial distribution using the simple fuzzy-sphere model.

\begin{figure}[htbp!]
    \centering
    \includegraphics[trim= 4.9cm 3.3cm 0.0cm 0.0cm, clip, width=0.8, width=0.49\textwidth]{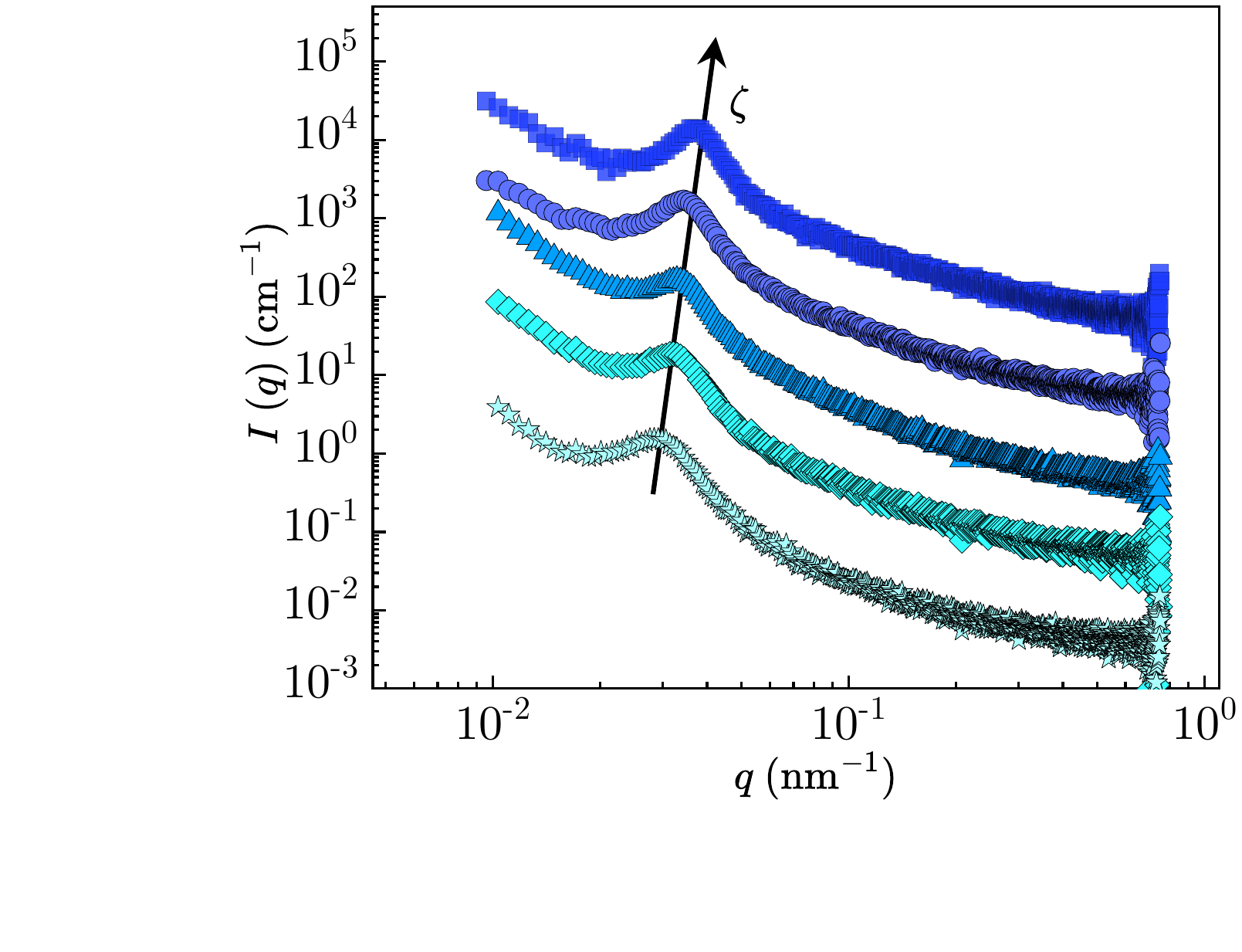}
    \caption{Small-angle X-ray scattering intensities $I(q)$ versus scattering vector $q$ of ULC microgels measured at 20~$^\circ$C at $0.04\pm 0.01$ (stars), $0.04\pm 0.01$ (diamonds), $0.04\pm 0.01$ (triangles),$0.04\pm 0.01$ (spheres), $0.04\pm 0.01$ (squares).
    The curves are shifted vertically by a factor of $10^1$ per $\zeta$, from low to high $\zeta$, for clarity.}
    \label{fig:SAXS_ULC}
\end{figure}

\section{Model for small-angle scattering fits}\label{sec:mod}

The fuzzy core-shell model consists of an interpenetrating layer of core and shell. 
The length of this layer is $2\sigma_{in}$. 
The outer surface is also characterized by a decreasing polymer density. 
The length of this region is $\sigma_{out}$. 
The widths of the inner and outer regions with constant density are $W_{core}$ and $W_{sh}$, respectively. 
For a particle with such a radial distribution, one can write the scattering amplitude $A(q)$ as:
\begin{eqnarray}
    A(q) &=& \Delta\rho_{sh}V_{sh}\Phi_{sh}(q,R_{out},\sigma_{out})\nonumber\\
    &+&(\Delta\rho_{core}-\Delta\rho_{sh})V_{core}\Phi_{core}(q,R_{in},\sigma_{in})
		\label{eq:mod1}
\end{eqnarray}
		
where $\Delta\rho$ is the difference between the scattering length density of the solvent and the core (or the shell). 
$V_{core}$ and $V_{sh}$ are the core and shell volumes, respectively. 
The radii are defined as $R_{in} = W_{core}+\sigma_{in}$, $R_{out}~=~W_{core}+2\sigma_{in}+ W_{sh}+\sigma_{out}$ and the total radius is $R=R_{in}+R_{out}$. 
The normalized Fourier transform of the radial density profile can be written as follows:

\begin{eqnarray}
	\Phi(q,R,\sigma)&= & \frac{1}{V_n} \Big[ \left( \frac{R}{\sigma^2}+\frac{1}{\sigma} \right)\frac{\cos[q(R+\sigma)]}{q^4}\nonumber\\ & + &\left( \frac{R}{\sigma^2} -\frac{1}{\sigma} \right) \frac{\cos[q(R-\sigma)]}{q^4}-\frac{3\sin[q(R-\sigma)]}{q^5\sigma^2} \nonumber\\ &-&\frac{2R\cos(qR)}{q^4\sigma^2}+\frac{6\sin(qR)}{q^5\sigma^2}\Big]  
\label{eq:mod2}
 	\end{eqnarray}
    
where $V_n=R^3/3+R\sigma^2/6$~\cite{Sco19}.

The model is then convoluted with a Gaussian to account for the particle size polydispersity $p$ and for the SANS data with the resolution function of the instrument.

Finally, a Lorentzian term is added to account for the high-q part of the data:

\begin{equation}
    I_L(q) = \frac{I_L(0)}{1+q^2\xi^2}\,,
\end{equation}

\noindent where $I_L(0)$ is the contribution of the Lorentzian term at $q = 0$, and $\xi$ is the average mesh size of the polymeric network, or in other words, the average distance between two crosslinked points.

\section{Viscosimetry}\label{sec:visc}

In suspensions of hard spheres, the viscosity increases with increasing sphere packing fraction \cite{Seg95}. 
In the limit of highly diluted samples, at low packing fractions, the course of $\eta_r$, the solution viscosity divided by the solvent viscosity, as a function of the sphere packing fraction $\phi$ is well described by the Einstein-Batchelor equation: 
\begin{equation}
    \eta_r = 1 + 2.5\phi +5.9\phi^2\, 
\end{equation} 

\begin{figure}[htbp!]
    \centering
    \includegraphics[trim= 5.5cm 2.9cm 0.0cm 0.0cm, clip, width=0.49\textwidth]{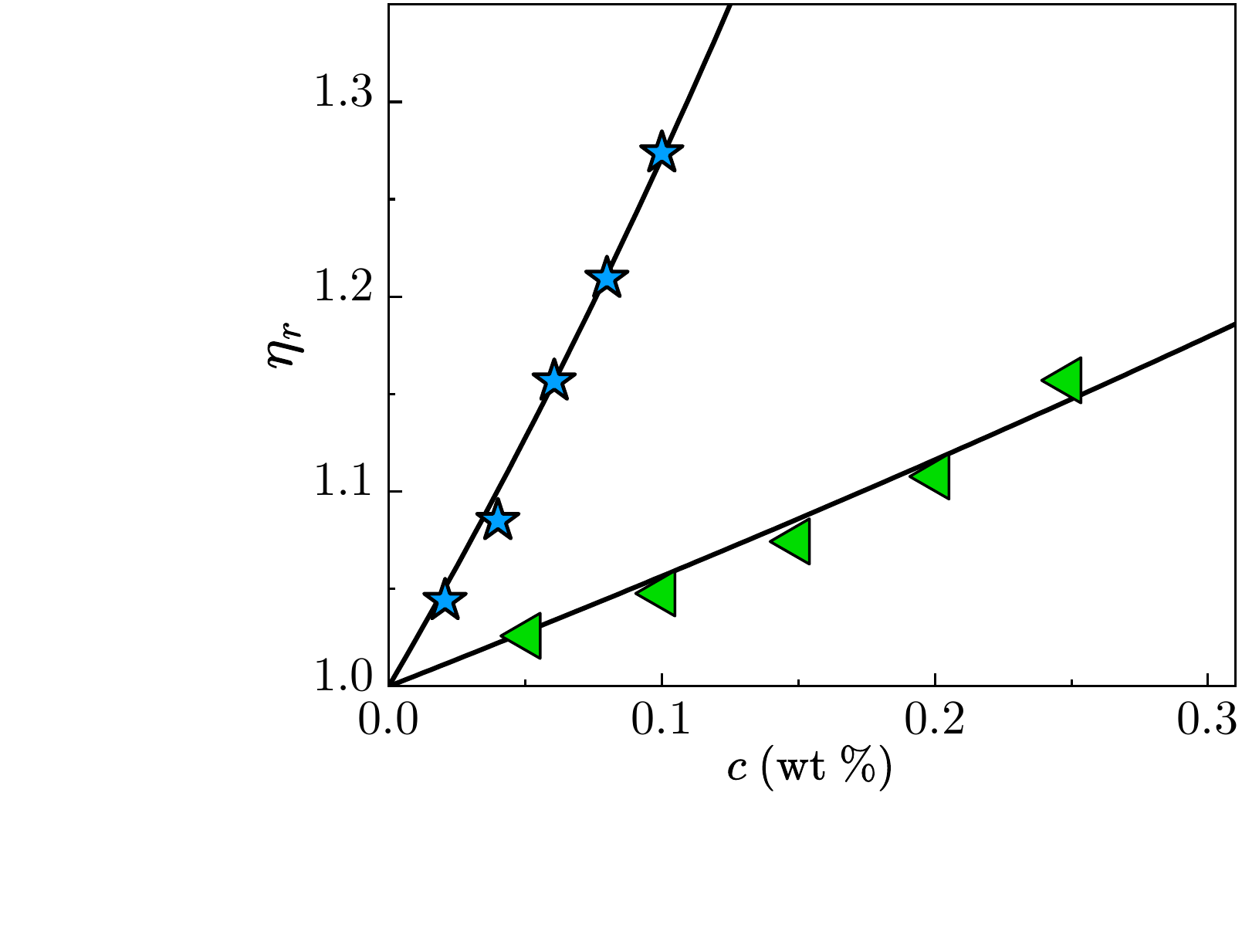}
    \caption{Relative viscosity $\eta_r$ of suspensions of RC microgels (triangles) and ULC microgels (stars) as a function of weight fraction $c$. Black solid lines are fits with Eq.~\ref{eq:EB}.}
    \label{fig:visc}
\end{figure}

This equation also describes the course of viscosity with increasing the concentration of microgels in solution, in the limit of highly diluted samples \cite{Bor99, Sen99}. 
The formal substitution of the real particle volume fraction $\phi$ with the generalized volume fraction $\zeta$, $\phi = \zeta$, is valid at low concentrations where the microgels do not experience any deswelling or deformation, allowing us to rewrite Eq.~\ref{eq:EB}:
\begin{equation}\label{eq:EB}
    \eta_r = 1 + 2.5\zeta +5.9\zeta^2\,.
\end{equation} 
The value of $\zeta$ is related to the microgel weight fraction in solution, $c$, by means of a multiplicative constant, $k$\cite{Gas13, Gas14, Sco16, Sco17, Sen99, Moh17, Con17, Con19, Pel16}. 
Substituting $\zeta = kc$ in the previous equation leads to $\eta_r = 1 + 2.5kc +5.9(kc)^2$. 
This relation is used to fit the data of the relative viscosity of the solution of RC and ULC shown in Fig.~\ref{fig:visc}. 
From these fits, the value of $k$ is determined and then used to obtain the values of $\zeta$ used in our study.

\section{Dynamic light scattering}\label{sec:DLS}

The DLS measurements were performed on a Mod3D-DLS Spectrometer (LS Instruments, Switzerland) equipped with a 660~nm Cobolt laser with a maximum power of 100~mW. 
5 mm cylindrical glass cells were used and placed in the temperature-controlled index-matching bath containing decalin. 
The scattered light was detected within a scattering angle range of 20$^\circ$ – 135$^\circ$ by avalanche photodiodes and processed by a LS Instrument correlator.

\begin{figure}[htbp!]
    \centering
    \includegraphics[trim= 12cm 3.5cm 11cm 0.0cm, clip, width=0.33\textwidth]{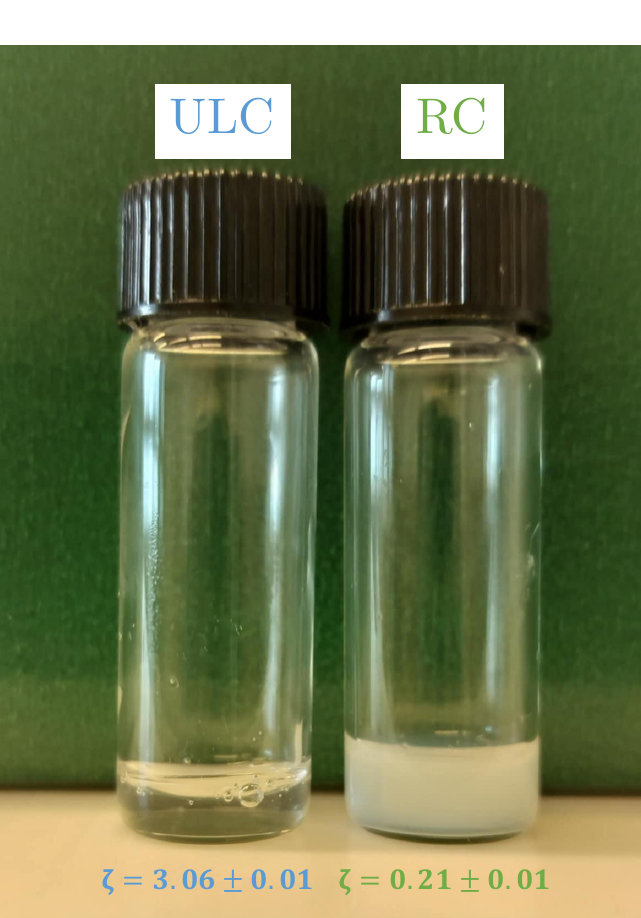}
    \caption{The image shows vials with suspensions of ULC (left) and RC microgels at $\zeta=3.06\pm~0.01$ and $\zeta=0.21\pm~0.01$, respectively.}
    \label{fig:photo}
\end{figure}

\begin{figure*}[htbp!]
    \centering
    \includegraphics[trim= 0cm 0cm 0.0cm 0.0cm, clip, width=0.8\textwidth]{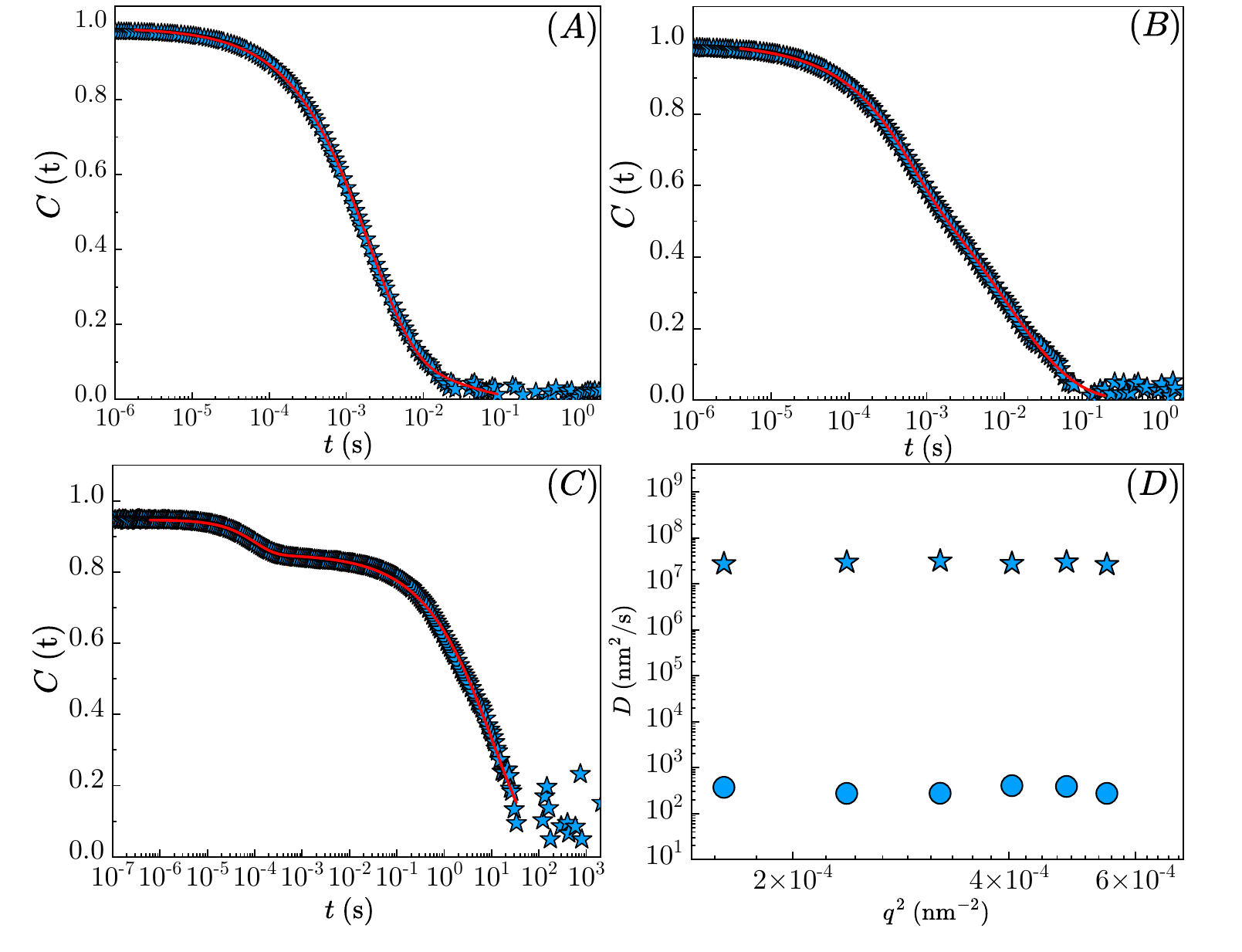}
    \caption{Intermediate scattering function (ISF) C(t) for ULC microgels at $q = 1.79 \times 10^{-2}~\mathrm{nm}^{-1}$, where the generalized packing fraction $( \zeta )$ ranges from (A)  $0.54\pm0.01$, (B) $0.82\pm0.01$ and (C) $3.06\pm0.01$. (D) short (circles) and long time (stars) diffusion (D ~(nm$^{2}$/s)) at $\zeta = 3.06\pm0.01 $ of ULC particles versus \( \ q^2 \).}
    \label{fig:DLS_SI}
\end{figure*}

\section{Small-angle neutron scattering}\label{sec:SANS}

The $q$ range of interest for the measurements performed with the D11 instrument at the Institut Laue-Langevin (ILL, Grenoble, France) was covered using three configurations: with sample-to-detector distances $d_\mathrm{SD} = 38.9$~m, $d_\mathrm{SD} = 10.2$~m and  $d_\mathrm{SD} = 2$~m, all with $\lambda = 0.524$~nm. The velocity selector operates with 10\% resolution. 
The instrument was equipped with a \(^3\)He detector with a pixel (8x4)mm.

To experimentally obtain the bulk modulus of the particle, we measured binary mixtures of a few microgels ($\zeta < 0.05$) and a high molecular weight ($M_n = 265,000$~g/mol) partially deuterated polyethylene glycol (d$_{83\%}$PEG). 
The concentration of the d$_{83\%}$PEG varied between 0 and $\approx 5~wt\%$.

\begin{figure*}[htbp!]
    \centering
    \includegraphics [width=0.8\textwidth]{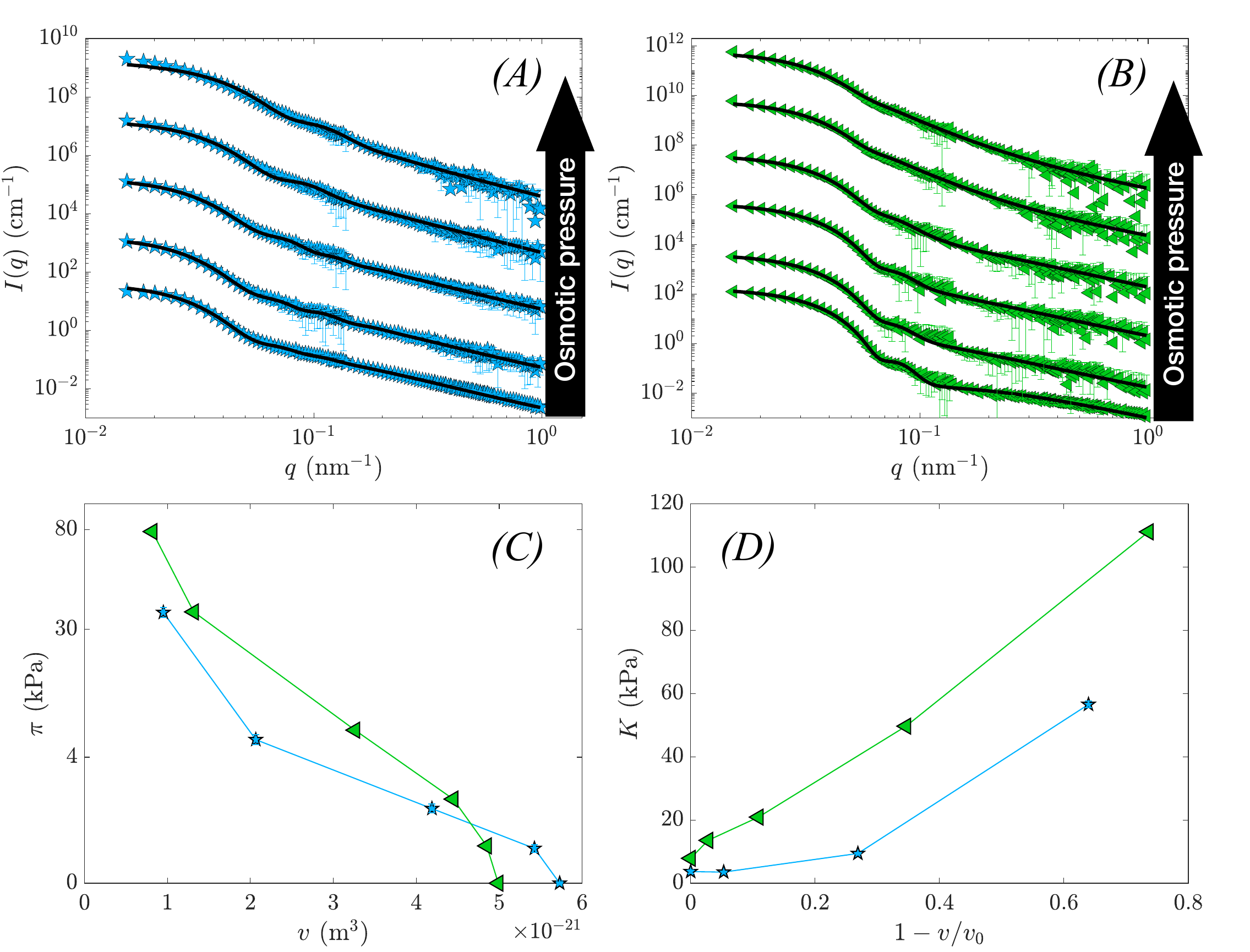}
    \caption{Small-angle neutron scattering intensities measured for the ULC (A) and RC microgels (B). 
    From bottom to top, the concentration of deuterated PEG in solution increases from 0 to $\simeq 5$~wt\%. 
    The bottom curves are therefore the form factor measure without PEG. 
    All measurements were performed at T =  20$^\circ$C (C) Osmotic pressure versus variation in volume of the microgels. 
    (D) Values of the bulk modulus versus particle compression.}
    \label{fig:K}
\end{figure*}

The neutron scattering length density of the d$_{83\%}$PEG is  $6.335 \cdot 10^{-6}$~\AA~$^{-2}$, i.e.~the same as D$_2$O.
As can be seen in Figs.~\ref{fig:K}(A) and \ref{fig:K}(B), the polymer is completely contrast matched when the suspension is prepared in pure D$_2$O and the only contribution to the coherent scattering comes from the form factor of the microgels.

The bottom curves in Figs.~\ref{fig:K}(A) and \ref{fig:K}(B) are measured without any d$_{83\%}$PEG, while the polymer concentration increases from bottom to top.
The solid lines are fits of the data with a model for a fuzzy core-shell structure \cite{Ber05, Ber06, Sco19MM} (see Sec.~\ref{sec:mod}).
From the fits, one can obtain the total size of the particles $R$ as a function of the wt\% of d$_{83\%}$PEG in suspension.
These values are used to compute the  volumes of the particles $v = \frac{4}{3}\pi R^3$ shown in Fig.~\ref{fig:K}(C).

The d$_{83\%}$PEG cannot penetrate the microgels since its hydrodynamic radius ($\approx$ 25~nm) is larger than the average microgel mesh-size $\xi$ \cite{Hou22, Sco22k, Hof24}.
This is also true in our case, as shown by the values of $\xi$ in Tab.~\ref{tab:SAXSPq}.
The effect of the d$_{83\%}$PEG is to exert an external osmotic pressure $\pi$ on the microgels that progressively deswell isotropically.
The relation between wt\% of d$_{83\%}$PEG  and $\pi$ has been determined experimentally \cite{Hou22} using a membrane osmometer.
This relation allows us to compute the values of $\pi$ in our suspensions, plotted in Fig.~\ref{fig:K}(C).

The bulk modulus is defined as:

\begin{equation}
    K = -v\frac{\partial \pi}{\partial v}\,,
\end{equation}

therefore, if we consider the slope between two consecutive points, $i$ and $i + 1$, in Fig.~\ref{fig:K}(C), we can obtain the value of $K$ for a given step of the compression as

\begin{equation}
    K_i = -v_i\frac{\pi_i - \pi_{i+1}}{v_i - v_{i+1}}\,.
\end{equation}

This is how we compute the values of $K$ plotted in Fig.~\ref{fig:K}(D) as a function of the microgel compression $1-v/v_0$, where $v_0$ is the volume of the microgel in the swollen state at 20$~^\circ$C measured without any d$_{83\%}$PEG in pure D$_2$O.
As can be seen, the bulk modulus of both microgels increases with increasing the compression in agreement with previous studies \cite{Hou22, Sco22k, Hof24}.
This means that the more a particle gets compressed, the harder it is to compress it further.

\section{Rheology}\label{sec:rheo}

Steady shear measurements were performed on a stress-controlled Anton Paar MCR301 rheometer using stainless steel cone and plate geometries (40 and 50~mm diameter, \( 1^\circ \) angle). 
The bottom plate is part of a Peltier unit, maintaining the temperature at  20\,\textdegree{}C with a solvent trap to minimize evaporation. 
No signs of wall slip, shear banding, or aging were observed at any packing fraction; typical characteristics of soft glasses~\cite{Bal08, besseling2010shear, meeker2004slip}.

\begin{figure*}[htbp!]
    \centering
    \includegraphics[trim= 0cm 8cm 0.0cm 0.0cm, clip, width=0.8\textwidth]{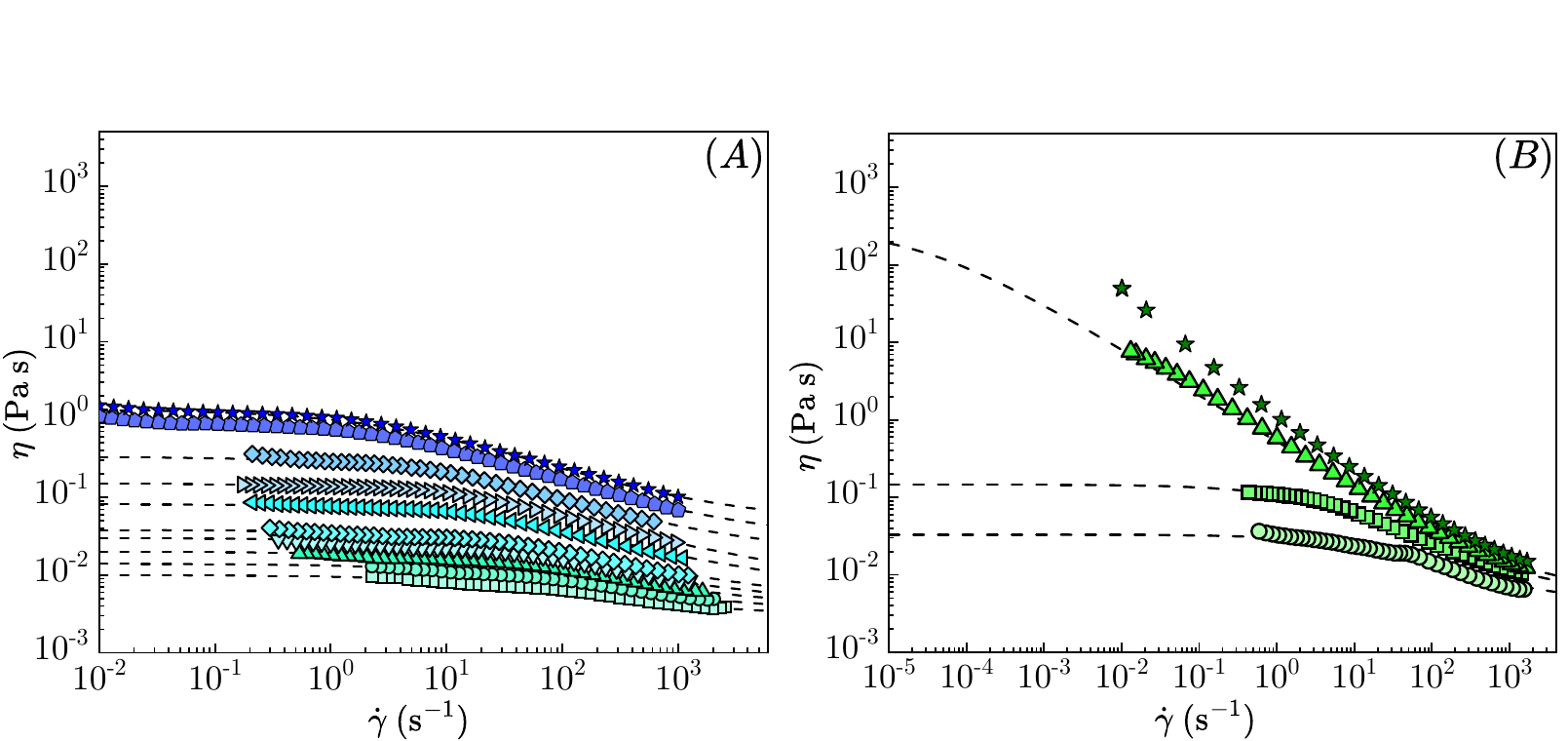}
    \caption{ Steady shear viscosity, \( \eta \), versus shear strain rate, \( \dot{\gamma} \), of (A) ULC suspensions with packing fractions from bottom to top: \( \zeta = 0.5 \) to 3.06 and (B) RC suspensions with packing fractions from bottom to top: \( \zeta = 0.6 \), to 0.7. 
    Dashed lines: fits of the data with Eq.~\ref{eq:Cross-equation}.}
    \label{fig:flow_curves}
\end{figure*}


 We derived the zero shear viscosity $\eta_0$ from steady shear measurements (Fig.~\ref{fig:flow_curves})  from the fit with Eq.~\ref{eq:Cross-equation}.
 
\begin{equation}\label{eq:Cross-equation}
    \frac{\eta - \eta_0}{\eta_0 - \eta_{\infty}} = \frac{1}{\left(1 + \frac{\gamma}{\gamma_c}\right)^m}\,,
\end{equation}

 \noindent where $m$ the power-law exponent counts for shear-thinning behavior, $\eta$  and $\eta_{\infty}$ the apparent and infinite shear viscosity, and $\gamma$ and $\gamma_c$ the strain and critical strain rate, respectively. 

\bibliographystyle{apsrev4-1}
\bibliography{References}